\documentclass[paper]{JHEP3}
\usepackage{amsmath}
\usepackage{amssymb}
\usepackage{cite}
\usepackage{epsfig}

\newcommand{\Imag}[1]{{\rm Im} \left\{ #1 \right\} }

\def\cO#1{{\mathcal{O}}\left(#1\right)}
\def\half{\tfrac12}

\def\diff{{\rm d}}

\def\QQ{Q\bar{Q}}
\def\ee{e^+e^-}

\def\as{\alpha_{\mbox{\scriptsize s}}}

\def\de{\delta}

\def\Om{\Omega}

\def\cN{{\mathcal N}}
\def\cI{{\mathcal I}}
\def\cE{{\mathcal E}}
\def\cC{{\mathcal C}}

\def\cH{{\mathcal H}}
\def\cT{{\mathcal T}}
\def\cM{{\mathcal M}}

\title{Exact solution of BFKL equation in jet-physics}
\author{G. Marchesini$^{\,a}$ {\it\small and\/} E. Onofri$^{\,b}$\\
  $a)$ Dipartimento di Fisica, Universit\`a di Milano-Bicocca and \\
  $\quad$ INFN, Sezione di Milano, Italy\\
  $b)$ Dipartimento di Fisica, Universit\`a di Parma and \\
  $\quad$ INFN, Gruppo Collegato di Parma, Italy }

\abstract{It has been recently found that the heavy quark-antiquark
  $\QQ$ pair multiplicity, in certain phase space region ($\QQ$ at
  short distance, soft and with small velocity), satisfies an
  evolution equation formally similar to the BFKL equation for the
  high energy scattering amplitude. We find the exact solution of the
  $\QQ$-equation and discuss the differences with the BFKL scattering
  amplitude}


\preprint{
  Bicocca-FT-04-3\\
  UPRF-2004-05\\
  \date{April 21, 2004} }

\begin{document}

\section{Introduction}\label{sec:introduction}

A way to reveal properties of QCD radiation consists in resumming
logarithmically enhanced terms in the perturbative expansion for
observables at short distances \cite{Basics} (logarithms in Feynman
diagrams come when vertices are in collinear or infrared
configurations).

Recently attention has been attracted \cite{DS} on short distance observables
for jet physics which have only infrared logarithms, to leading order. An
observable of this type, considered in \cite{DS}, is the distribution in $E_{\rm
  out}$, the energy deposited by QCD radiation into a region away from all jets.
Here one resums (to leading order) powers in $\as\ln Q/E_{\rm out}$ with $\as$
the QCD coupling and $Q$ the hard momentum scale (e.g. the center of mass energy
in $\ee$ annihilation).  Another observable, considered in \cite{MM}, is the
multiplicity of a heavy quark-antiquark system $\QQ$ of mass $\cM$ in a phase
space region in which collinear singularities are absent. Here one resums (to
leading order) powers in $\as\ln Q/\cM$.  These logarithms originate, in the
parton language, from successive gluon branching\footnote{These type of
  logarithmic terms are present also in all non-global jet-shape observables
  which have also collinear singularities such as the Sterman-Weinberg
  distribution \cite{SW}.  Here their contribution is beyond leading logarithmic
  order.}  and are strictly related to the non-Abelian structure of QCD
vertices.  The resummation of these ``non-Abelian'' logarithms raises various
surprises.

The first surprise is that standard collinear factorization methods
\cite{Basics} do not work here \cite{DS} and these non-Abelian
logarithms cannot be resummed simply into Sudakov factors.  The way
used to resum them was to use the distributions for multi-soft gluons
emitted in any angular configuration.  They are known \cite{BCM}, but
only in the planar approximation in $SU(N_c)$.  In the case of the
$E_{\rm out}$-distribution, the resummation is performed by numerical
methods \cite{DS} or by a non-linear evolution equation \cite{BMS}
which reflects the non-Abelian structure of the QCD branching. In the
case of the $\QQ$-multiplicity, the resummation is performed \cite{MM}
by a linear equation.

The second surprise is that the equations resumming these non-Abelian
logarithms are similar to the equations obtained in high energy
physics in which one resums leading logarithms in the energy
(rapidity).  In particular, the equation for the $\QQ$-multiplicity is
similar to the Balitski-Fadin-Kuraev-Lipatov (BFKL) equation
\cite{BFKL} for the high-energy elastic scattering amplitude. However,
this similarity is only {\it formal} since the relevant momentum
configurations of the soft gluon ensemble are completely different.
For the $\QQ$-multiplicity all gluon angles, $\theta_{ij}$, are of
comparable order while their transverse momenta, $q_{ti}$, are
strongly ordered.  In the high energy case the opposite is true:
comparable $q_{ti}$ and strongly ordered $\theta_{ij}$. In \cite{MM}
one finds the discussion of this point.

The differences in the relevant phase space for of the soft gluon
ensemble have important consequences.
First of all the QCD coupling enters differently the two equations.
This is due to the fact that the arguments of the QCD running coupling
for the soft gluons are given by $q_{ti}$ which are ordered ($\QQ$
case) or comparable (BFKL case).  As a result the expansion parameter
for the leading order distributions is $\tau\!\sim\!\as(Q)\ln Q/\cM$
in the $\QQ$ case and $\tau\!\sim\!\as\,Y$ in the BFKL case, with
$\as$ the fixed coupling and $Y$ the rapidity (logarithm of energy).
This implies that the asymptotic regime for large $\tau$ is physically
accessible for the high energy case while it is quite extreme for the
$\QQ$ case.

A second difference is that the variables of comparable order are
dimensionful ($q_{ti}$ in the BFKL case) or dimensionless
($\theta_{ij}$ for the $\QQ$ case).  Therefore, while $q_{ti}$ are
unbounded (total energy much larger than $q_{ti}$), angles are
obviously bounded.  As a result the BFKL equation is equivalent to a
diffusion equation in a translational invariant system (using the
logarithm of impact parameter), while for the $\QQ$-equation there is
a boundary and translational invariance is lost.
Only in infinitely boosted frames one could try to neglect the angular
limitation (see \cite{MM}) so that the two equations would become
exactly the same. In this case one could use all results of high
energy studies \cite{smallx} to describe the $\QQ$-multiplicity.

In this paper we directly study and explicitly solve the
$\QQ$-multiplicity equation. We do not need to use a boosted frame, we
work in a general frame including the $\ee$ center of mass.  Moreover
the exact solution is valid for any value of the expansion parameter
$\tau\!\sim\!\as(Q)\ln Q/\cM$ not only in the asymptotic regime.  We
compare this solution with the BFKL high energy amplitude. We find
that the effect of the finite range in $\theta_{ij}$ is relevant even
at large $\tau$. Here both distributions are dominated by the Pomeron
intercept but have different prefactors.  It could be considered as a
third surprise that the mathematical structure of this equation is
quite rich and deserves a special study \cite{DPFO}.

In section \ref{sec:equations} we describe the observable
($\QQ$-multiplicity) and report the evolution equation in \cite{MM}
resumming leading logarithms.  In section \ref{sec:BFKL} we discuss
the difference between the $\QQ$-multiplicity and BFKL equations.
With the aim to learn the strategy to discuss the solution of the
$\QQ$-equation, in section \ref{sec:model} we introduce and study a
simple solvable model which bears features which are similar to the
$\QQ$-multiplicity equation.  In section \ref{sec:solution} we present
the exact solution of the $\QQ$-equation (details are given in
Appendix \ref{app:solution}).  In section \ref{sec:conclusion} we
summarize the properties of the solution.

\section{The observable and the equation}\label{sec:equations}

The observable introduced in \cite{MM} is the multiplicity
distribution of heavy quark-antiquark system $\QQ$ of mass $\cM$ and
momentum $|\vec{k}|$ produced in $e^+e^-$ annihilation with center of
mass energy $Q$ (similar observable could equally well be defined for
deep inelastic scattering or hard events in hadron-hadron collisions).
The distribution was studied for $Q\!\gg\!\cM$ where the perturbative
coefficients are enhanced by powers of $\ln Q/\cM$.
Moreover the $\QQ$ system was considered near threshold
($\cM^2\!-\!4M^2\!\sim\!M^2$ with $M$ the heavy quark mass) and with
small velocity $v=|\vec{k}|/E_k$ so that there are no collinear
singularities.
As a consequence, the leading logarithmic contributions
($\as^{2+n}\,\ln^n Q/\cM$) are obtained by considering soft secondary
gluons emitted in all angular region off the primary quark-antiquark.
The $\QQ$ system results from the decay of one of these soft gluons
(actually the softest one).

As shown in \cite{MM}, to leading logarithmic order, the
$\QQ$-multiplicity distribution factorizes into the inclusive
distribution $I$ for the emission of the soft off-shell gluon of mass
$\cM$ and momentum $|\vec{k}|$ and the distribution for its successive
decay into the $\QQ$ system
\begin{equation}
  \label{eq:origine}
\frac{E_k\,dN}{d\cM^2\,d|\vec{k}|}=\frac{\as^2\,C_F}{3\pi^2\cM^2}
\sqrt{\frac{\cM^2-4M^2}{\cM^2}}\frac{\cM^2+2M^2}{\cM^2}\cdot I\,,
\end{equation}
with $I$ a function of $\as\ln{Q}/{\cM}$ and of $v$ (we work in the
$\ee$ center of mass).  Instead of $\as\ln{Q}/{\cM}$, it is convenient
to introduce the equivalent variable\footnote{To reconstruct the
  argument of $\as$ one needs to go beyond leading order in the soft
  limit.}
\begin{equation}
  \label{eq:tau}
  \tau\>=\>
\int_{\cM}^{Q} \frac{dq_t}{q_t}\frac{N_c\,\as(q_t)}{\pi}=
\frac{2N_c^2}{11N_c-2n_f}
\ln\left(\frac{\ln Q/\Lambda}{\ln \cM/\Lambda}\right)\,,
\end{equation}
with $N_c$ the number of colours, $n_f$ the number of quark flavours
and $\Lambda$ the QCD scale.

We consider the $\ee$ annihilation in the center of mass so that, for
soft secondary radiation, the primary quark-antiquark are
back-to-back. However, to resum the leading logarithmic contributions
(via use of recurrence relations), one needs to consider a general
case in which soft gluons are emitted off a hard dipole forming an
angle $\theta$. The inclusive distribution for a soft off-shell gluon
of mass $\cM$ and velocity $v$ emitted off a dipole forming an angle
$\theta$ will be denoted by
\begin{equation}
  \label{eq:rho}
  I(\rho,\tau,v)\,,\qquad \rho=\half(1\!-\!\cos\theta)\,.
\end{equation}
The physical case corresponds to $\theta\!=\!\pi$.  For $Q\!\sim\!\cM$
the inclusive distribution is the Born contribution given by
($\tau\!=\!0$)
\begin{equation}
  \label{eq:Born}
  I(\rho,0,v)=\int\frac{d\Om_k}{4\pi}\frac{v^2(1-\cos\theta_{ab})}
{(1-v\cos\theta_{ak})(1-v\cos\theta_{kb})}=\rho\,f(\rho,v)\,,
\end{equation}
with $\rho=\half(1\!-\!\cos\theta_{ab})$. For $\cM\!<\!Q$ secondary
radiation starts to contribute.

For $v\!=\!1$ there is a collinear logarithmic divergence in
\eqref{eq:Born} which needs to be regularized and resummation of
secondary radiation leads to the standard multiplicity function
\cite{Basics}, a distribution with both collinear and soft logarithms.
Here we have $v\!<\!1$ and only soft logarithms are involved.

To resum the secondary radiation to leading order one needs the
multi-soft gluon emission distribution in all angular configurations
as given in \cite{BCM}.  This distribution can be represented as a
successive branching in which a general dipole of massless momenta
$q_i$ and $q_j$ emits a soft massless gluon $q$
\begin{equation}
\label{eq:branch}
dw_{ij}(q)=\frac{N_c\,\as(q_t)\,dq_t}{\pi\,q_t}\cdot
\frac{d\Omega_q}{4\pi}\frac{2\,\rho_{ij}}{\rho_{iq}\,\rho_{qj}}=
d\tau\cdot 
\frac{d\Omega_q}{2\pi}\frac{\rho_{ij}}{\rho_{iq}\,\rho_{qj}}\,.
\end{equation}
From this branching distribution, taking into account (strong) energy
ordering, one deduces (see \cite{MM}) the following evolution equation
resumming soft logarithms (neglect $\tau$ and $v$ dependence)
\begin{equation}
\begin{split}
\partial_{\tau}\,I(\rho_{ab},\tau)
=\int\frac{d\Omega_q}{2\pi}\frac{\rho_{ab}}{\rho_{aq}\,\rho_{qb}}
[I(\rho_{aq},\tau)+I(\rho_{qb},\tau)-I(\rho_{ab},\tau)]\,,
\end{split}
\end{equation}
with $\rho_{ij}=\half(1\!-\!\cos\theta_{ij})$.  The $v$ dependence,
here understood, is coming from the initial condition \eqref{eq:Born}
at $\tau\!=\!0$.  The first two terms in the integral originate from
emission out of $aq$ and $qb$ dipoles which result from the branching
of the parent dipole $ab$.  The third term originates from virtual
corrections in Feynman diagrams.

Since the off-shell gluon decaying into the $\QQ$ system is the
softest one, the initial condition at $\tau\!=\!0$ is given by the
Born distribution \eqref{eq:Born} which vanishes for
$\rho_{ab}\!\to\!0$. This implies that $I(\rho_{ab},\tau,v)$ vanishes
for $\rho_{ab}\!\to\!0$ and that the kernel is regular for
$\rho_{aq}\!\to\!0$ or $\rho_{qb}\!\to\!0$.  To make explicit the
regularity of the kernel we use the splitting (for simplicity we
replace $\rho_{ab}\!\to\!\rho$, $\rho_{aq}\!\to\!\rho_1$ and
$\rho_{qb}\!\to\!\rho_2$)
\begin{equation}
  \frac{\rho}{\rho_1\,\rho_2}=\frac{1}{\rho_1}+\frac{1}{\rho_2}+
  \frac{\rho-\rho_1-\rho_2}{\rho_1\,\rho_2}\,,
\end{equation}
where the last term is regular both for $\rho_1\!\to\!0$ and
$\rho_2\!\to\!0$.  The evolution equation becomes
\begin{equation}
\partial_{\tau}\,I(\rho,\tau)=\int\frac{d\Omega_q}{2\pi}\left\{
\frac{2}{\rho_2}\left[\frac{\rho}{\rho_1}\,I(\rho_1,\tau)-I(\rho,\tau)\right]-
  \frac{\rho-\rho_1-\rho_2}{2\,\rho_1\,\rho_2}\,I(\rho,\tau)\right\}.
\end{equation}
Using 
%\begin{equation}
%  \int_0^{2\pi}\frac{d\phi}{2\pi(a-b\cos\phi)}=
%\frac{1}{\sqrt{a^2-b^2}}\,,
%\end{equation}
%one finds
\begin{equation}
\int\frac{d\Omega_q}{2\pi}\>\frac{2}{\rho_2}=
\int_0^{1}\frac{d\rho_1}{|\rho_1-\rho|}\,,
\qquad
\int\frac{d\Omega_q}{2\pi}\>\frac{\rho-\rho_1-\rho_2}{\rho_1\,\rho_2}=
-\int_{\rho}^{1}\frac{d\rho_1}{\rho_1}\,,
\end{equation}
one finds
\begin{equation}
\partial_{\tau}\,I(\rho,\tau)=\int_0^1\frac{d\rho_1}{|\rho_1-\rho|}
\left[\frac{\rho}{\rho_1}I(\rho_1,\tau)-I(\rho,\tau)\right]
+\int_{\rho}^1\frac{d\rho_1}{\rho_1}\,I(\rho,\tau)\,.
\end{equation}
Using \mbox{$\rho_1=\eta\rho$} in the range \mbox{$0<\rho_1<\rho$} and
\mbox{$\rho_1=\rho/\eta$} in the range $\rho<\rho_1<1$, one gets the
final form of the evolution equation for the inclusive $\QQ$
distribution resumming all leading soft logarithms
\begin{equation}
  \label{eq:eveqMM}
  \partial_{\tau}I(\rho,\tau)=\int_0^1\frac{d\eta}{1-\eta}
\left(\eta^{-1}I(\eta\rho,\tau)-I(\rho,\tau)\right)
+\int_{\rho}^1\frac{d\eta}{1-\eta}
\left(I(\eta^{-1}\rho,\tau)-I(\rho,\tau)\right)\,.
\end{equation}
The lower limit $\eta\!>\!\rho$ in the second integral ensures that
the argument of $I(\rho/\eta,\tau)$ remains within the physical region
$\rho/\eta<1$. 

In \cite{MM} the solution of \eqref{eq:eveqMM} was considered only for
$\rho\ll 1$. This corresponds to taking the $\ee$ system in a boosted
frame however it was not clear how to go back into a non-boosted
frame. The reason for this limit is that it was argued that
\eqref{eq:eveqMM} reduces to the BFKL equation. Indeed, if one
neglects the lower limit $\eta\!>\!\rho$ in the second integral of
\eqref{eq:eveqMM}, the equation is formally the same as the BFKL
equation
\begin{equation}
  \label{eq:eveqBFKL}
  \partial_{\tau}T(\rho,\tau)=\int_0^1\frac{d\eta}{1-\eta}
\left(\eta^{-1}T(\eta \rho,\tau)-T(\rho,\tau)\right)
+\int_{0}^1\frac{d\eta}{1-\eta}
\left(T(\eta^{-1}\rho,\tau)-T(\rho,\tau)\right),
\end{equation}
where $T(\rho,\tau)$ is the high energy elastic scattering amplitude
in the impact parameter representation, $\rho=b^2$ is the square of
the impact parameter and $\tau=N_c\as/\pi\,Y$ (here $\as$ is the fixed
QCD coupling and $Y$ the rapidity).
 
Neglecting for $\rho\ll1$ the lower limit in the second integral of
\eqref{eq:eveqMM} one can use well known results in high energy
physics such as the Pomeron intercept $\alpha_P =\chi(0)\,N_c\as/\pi$
where
\begin{equation}
  \label{eq:chi}
  \chi(k)=2\psi(1)-\psi(\half+ik)-\psi(\half-ik)\,,\qquad
\chi(0)=4\ln2\,,
\end{equation}
is the BFKL characteristic function. In this way one determines the
asymptotic behaviour for large $\tau$ of the inclusive distribution
\begin{equation}
  \label{eq:asymMM}
  I(\rho,\tau)\sim\frac{e^{\chi(0)\,\tau}}{\sqrt{\tau}}\,.
\end{equation}
Neglecting the lower bound in the second integral in \eqref{eq:eveqMM}
implies that the inclusive distribution $I(\rho,\tau)$ needs to be
defined in the full range $\rho\!>\!0$ including the region
$\rho\!>\!1$ outside the physical range.
The extension to the non-physical region $\rho\!>\!1$ seems not
crucial since at high $\tau$ the distribution rapidly decreases at
large $\rho$ so that one expects that such an extension should not
give a significant contribution to the second integral of
\eqref{eq:eveqMM}. A second argument supporting this expectation is
obtained by observing that the $\QQ$-equation is invariant if we
change $\rho\!\to\!a\rho$ and by sending $a\!\to\!0$ one could neglect
the lower bound in the second integral of \eqref{eq:eveqMM}.
It was however pointed out by Gavin Salam that a cutoff in $\rho$
actually gives even asymptotically an effect and generates \cite{MS}
an additional factor $\tau^{-1}$ in the large $\tau$ behaviour in
\eqref{eq:asymMM}. This fact is discussed in the following.

\section{Comparison with BFKL equation}\label{sec:BFKL}

Let us first recall the features of the solution to the BFKL equation
which are relevant for the discussion of the solution of
\eqref{eq:eveqMM}.
The impact parameter $b$ in the elastic scattering amplitude at high
energy $T(b^2,\tau)$ is taken in the full range $b\!>\!0$. In terms of
the function
\begin{equation}
  \label{eq:phiBFKL}
\phi_0(x,\tau)=e^{x/2}\,e^{-4\ln2\;\tau}\,T(b^2,\tau)\,,
\qquad b^2=e^{-x}\,,\qquad -\infty<x<\infty\,,
\end{equation}
the BFKL equation \eqref{eq:eveqBFKL} can be written in the form
\begin{equation}
  \label{eq:H-BFKL}
\partial_\tau\phi_0(x,\tau)=
\int_{-\infty}^\infty\,dy\,
\frac{\phi_0(y,\tau)-\phi_0(x,\tau)}{2\sinh\half|y-x|}
=-H_0\cdot\phi_0\,.
\end{equation}
The kernel is translation invariant, hence diagonal in momentum space
\begin{equation}
  \label{eq:E}
\begin{split}
&H_0\cdot e^{ikx}=
-\int_{-\infty}^\infty\,dy\,\frac{e^{iky}-e^{ikx}}{2\sinh\half|y-x|}
=E_0(k)\,e^{ikx}\,,\\
&E_0(k)=4\log 2-\chi(k) =14\,\zeta(3)\,k^2+\cdots
\end{split}
\end{equation}
The BFKL equation is then equivalent to a Euclidean Schroedinger
equation (a diffusion equation) for a ``free'' particle with energy
(dispersion relation) $E_0(k)$ and the solution is given by a wave
packet
\begin{equation}
  \label{eq:phi0}
 \phi_0(x,\tau)=\int_{-\infty}^{\infty}\frac{dk}{2\pi}
\,e^{ikx}\,C_0(k)\,e^{-\tau\,E_0(k)}\,,
\end{equation}
where $C_0(k)$ is defined in terms of the initial condition.
At large $\tau$ the distribution is dominated by the small $k$ region
and consequently the solution is what we expect for a diffusion at
large times, i.e.  typically a Gaussian with standard deviation
$\propto \tau$
\begin{equation}
  \label{eq:Gauss}
  \phi_0(x,\tau)\simeq C_0(0)\,
  \frac{\exp\{-\frac{(x-x_0)^2}{2D\tau}\}}{\sqrt{2\pi D\,\tau}}\,,
\qquad C_0(0)\,= \int_{-\infty}^{\infty}dx\,\phi_0(x,0)\,.
\end{equation}
with $D=28\,\zeta(3)$. Translational invariance implies that the
integral of the distribution is conserved
\begin{equation}
  \label{eq:consBFKL}
\frac{\diff}{\diff\tau}\int_{-\infty}^{\infty}dx\,\phi_0(x,\tau)=0\,,
\end{equation}
which is obviously the case also for the asymptotic expression
\eqref{eq:Gauss}.

We come now to discuss the $\QQ$-equation.  As in \eqref{eq:phiBFKL},
in the jet physics case we introduce the function
\begin{equation}
\phi(x,\tau)=e^{x/2}\,e^{-4\ln2\;\tau}\,I(\rho,\tau)\,,
\>\>\qquad \rho=e^{-x}\,,\qquad 0<x<\infty\,,
\end{equation}
and \eqref{eq:eveqMM} becomes
\begin{equation}
  \label{eq:H-MM}
\partial_\tau\phi(x,\tau)\>=
\int_0^\infty\,dy\,\frac{\phi(y,\tau)-\phi(x,\tau)}{2\sinh\half|y-x|}
\>\>-\>2\log(1+e^{-x/2})\,\phi(x,\tau)=-H\cdot\phi\,.
\end{equation}
To discuss the solution of \eqref{eq:H-MM} we can still conveniently
follow the analogy with the Schroedinger equation, but we have to take
into account two different features: the existence of a boundary at
$x=0$ and a short range repulsive ``potential''
$V(x)=2\log(1+e^{-x/2})$ which is responsible for the decrease in time
of the integrated density
\begin{equation}
  \label{eq:PhiMM}
\frac{\diff}{\diff\tau}\int_0^{\infty}dx\,\phi(x,\tau)=
-2\int_0^{\infty}dx\ln(1+e^{-x/2})\,\phi(x,\tau)<0\,.
\end{equation}
(If we prefer to think in terms of diffusion processes, $V$ is
connected to a probability density for killing the particle).  At
large $x$ (i.e. small $\rho$) the effect of the boundary and the
potential become (almost) negligible and the particle is essentially
free with energy $E_0(k)$ in \eqref{eq:E}. It will be natural to
discuss the problem in terms of stationary waves, which should be
given, at least for large $x$, by a shifted wave $\sin(kx+\delta(k))$,
as is well--known from elementary wave mechanics.

In order to explore this picture without technical complications, we
shall first study, in the next section, a solvable model with and
without boundary at $x=0$.

\section{A simplified solvable model}\label{sec:model}
We consider the following evolution equation which bears some features
similar to inclusive jet physics cases in \eqref{eq:eveqMM}
\begin{equation}
  \label{eq:sim-MM}
\partial_{\tau}\cI(\rho,\tau)=\int_0^1d\eta\,\eta^{-1}\cI(\eta\rho)
+\int_{\rho}^1d\eta\,\cI(\eta^{-1}\rho)\,.
\end{equation}
This is obtained from \eqref{eq:eveqMM} by neglecting the singularity
at $\eta=1$ and, correspondingly, neglecting the regularization due to
the virtual correction. The crucial point of the discussion here is
the presence of the cutoff $\rho\!<\!1$ which is reflected in the
lower bound in the second integral.

To discuss the effect of the cutoff $\rho\!<\!1$ we consider the
associated evolution equation (which bears some resemblance with the
BFKL equation) for a function $\cT(\rho,\tau)$ with $\rho$ in the full
range $\rho\!>\!0$
\begin{equation}
  \label{eq:sim-BFKL}
\partial_{\tau}\cT(\rho,\tau)=\int_0^1d\eta\,\eta^{-1}\cT(\eta \rho)
+\int_{0}^1\,\diff\eta\,\cT(\eta^{-1}\rho)\,.
\end{equation}
Introducing as in \eqref{eq:phiBFKL} the function
\begin{equation}
  \label{eq:cT}
\psi_0(x,\tau)= e^{x/2}\,e^{-4\tau}\,\cT(\rho,\tau)\,,
\qquad \rho=e^{-x}\,,\qquad -\infty<x<\infty\,,
\end{equation}
we obtain ($4\ln2$ is replaced by $4$ so that the integrated
distribution of $\psi_0$ is conserved)
\begin{equation}
  \label{eq:sBFKL}
\partial_\tau\psi_0(x,\tau)=
\int_{-\infty}^\infty\,dy\,\,e^{-\half|y-x|}\,
(\psi_0(y,\tau)-\psi_0(x,\tau))\>=\>-\cH_0\cdot\psi_0\,.
\end{equation}
Due to translation invariance the evolution equation is diagonal in
momentum space and the dispersion relation is now
\begin{equation}
  \label{eq:cE}
\cH\cdot e^{ikx}=\cE_0(k)\,e^{ikx}\,,\qquad  
\cE_0(k)=\frac{16k^2}{(1+4\,k^2)}  
\end{equation}
thus giving the asymptotic solution \eqref{eq:Gauss} with $D=32$.

Consider now Eq.~\eqref{eq:sim-MM}. We introduce the function
\begin{equation}
  \label{eq:cI}
\psi(x,\tau)=e^{x/2}\,e^{-4\,\tau}\,\cI(\rho,\tau)\,,
\>\>\qquad \rho=e^{-x}\,,\qquad 0<x<\infty\,,
\end{equation}
and obtain
\begin{equation}
  \label{eq:sMM}
\partial_\tau\psi(x,\tau)\>=
\int_{0}^\infty\,dy\,\,e^{-\half|y-x|}\,(\psi(y,\tau)-\psi(x,\tau))\>-
\>2\,e^{-\frac{x}{2}}\,\psi(x,\tau)\>=\>-\cH\cdot\psi\,.
\end{equation}
Since at large $x$ the effect of boundary and potential
$V(x)=2e^{-x/2}$ become negligible, the particle is essentially free
with energy $\cE_0(k)$. At large $x$ the solution should be given by a
shifted wave $\sin(kx+\delta(k))$.  This model is actually solvable
and this expectation turns out to be correct.

To solve the model, observe that the kernel of \eqref{eq:sMM} is
easily identified with the Green's function of the operator
$-\diff/\diff x^2$ with boundary condition at $x=0$
\begin{equation}
  \label{eq:bc}
  \partial_x \psi(0,t)=\half \psi(0,t)\,.
\end{equation}
This implies that the a complete set of eigenfunctions is given by 
\begin{equation}
  \label{eq:u}
  u(x,k)\>=\>\frac{\sin(kx)+ 2k\,\cos(kx)}{\sqrt{1+4k^2}}\>=\>
\sin(kx+\de(k))\,,\qquad \de(k)=\tan^{-1} 2k \,.
\end{equation}
with the same eigenvalue $\cE_0(k)$ of (\ref{eq:cE}). The
solution is then
\begin{equation}
  \label{eq:psi}
\psi(x,\tau)=\sqrt{\frac{2}{\pi}}\int_{0}^{\infty}\,dk
\,u(x,k)\,\cC(k)\,e^{-\tau\,\cE_0(k)}\,,
\qquad
  \cC(k)=\sqrt{\frac{2}{\pi}}\int_0^{\infty} dx\,u(x,k)\,\psi(x,0)\,.
\end{equation}
For instance, by taking $\cI(\rho,0)=\rho$ (see \eqref{eq:Born}) one
has $\cC(k)= 8k/(1+4k^2)^{3/2}$.  The behavior of the solution at
large $\tau$ is dominated by the small $k$ behavior of the integrand.
Because of the presence of the boundary the eigenfunctions vanish
linearly at $k=0$, also the amplitude $\cC(k)$ vanishes linearly
\begin{equation}
  \label{eq:CK}
  \cC(k) = c_0 k\left(1 + \cO{k^{2}}\right),
\end{equation}
and then for  large $\tau$
\begin{equation}
\label{eq:psi-asy}
  \psi(x,\tau) \simeq c_0
  \frac{(x+2)e^{-\frac{x^2}{2D\tau}}}{D\tau\,\sqrt{2\pi D\tau}}
\end{equation}
with $D=32$. The prefactor \mbox{$x$+$2$} ensures that the asymptotic
solution satisfies the boundary condition (\ref{eq:bc}).  The solution
at large $\tau$ has an additional factor $(D\tau)^{-1}$ w.r.t. the
free evolution case in (\ref{eq:Gauss}) (see also \cite{MS}).

\section{Solution of the jet equation}\label{sec:solution}

The discussion on the solution of the complete problem for the
inclusive off-shell gluon distribution $I(\rho,\tau)$ will follow the
same path as in the model case.  We have to take into account the
effect of potential $V(x)$ and the presence of the border at $x=0$.
Since the potential is short range, for a function with support far
from the boundary, the operator $H$ in \eqref{eq:H-MM} is
indistinguishable from the BFKL one in \eqref{eq:H-BFKL}; it follows,
like in ordinary quantum mechanics, that the continuous spectrum is
unaffected. Also, since the potential is repulsive, we do not even
have bound states. On the basis of what we have learned from the
solvable model, we expect that the eigenfunctions $u(x,k)$ are given
by
\begin{equation}
\label{eq:UMM}
\begin{split}
    &H\cdot u(x,k)= E_0(k)\,u(x,k)\\
    &u(x,k)\simeq \sin(kx+\de(k))\,,\qquad x\gg1\,,
\end{split}
\end{equation}
with $E_0(k)$ given by the BFKL dispersion relation \eqref{eq:E}.

To find the expression for the phase shift $\de(k)$ and the
eigenfunctions at finite $x$ may seem problematic since, unlike the
simplified model, there does not exist any simple boundary condition
to be imposed to the eigenstates.  Both the phase shift $\de(k)$ and
the eigenfunctions $u(x,k)$ however can be determined by an expansion
method which can be pushed to any order. Here we outline the method
which will be reported in detail in Appendix~\ref{app:solution}.

One starts by defining the eigenfunction at large $x$ as the shifted
waves in \eqref{eq:UMM}. A first estimate of the phase shift is then
obtained by requiring that the correction at large $x$ is
exponentially small.  This also requires to determine the
subasymptotic correction to the eigenfunction.  Applying the operator
$H$ in \eqref{eq:H-MM} to a plane wave we get
\begin{eqnarray*}
  H\,e^{ikx} &=& e^{ikx}\, \int_0^\infty
\frac{e^{ik(y-x)}-1}{2\sinh\half|x-y|}\,\diff y\, - 
2 \log(1+e^{-x/2})\,e^{ikx}\\\nonumber
&=& e^{ikx}\,\int_{-\infty}^\infty
\frac{e^{ik(y-x)}-1}{2\sinh\half|x-y|}\,\diff y -
 e^{ikx}\,\int_{x}^\infty\frac{e^{-iky}-1}{2\sinh\half y}\,\diff y -
2 \log(1+e^{-x/2})\,e^{ikx}\\ \nonumber
&=& E_0(k)\,e^{ikx}  - \frac2{1+2ik}e^{-x/2}  + \cO{e^{-x}}
\end{eqnarray*}
The leading term $\cO{e^{-x/2}}$ can be canceled by taking a
superposition of $e^{ikx}$ and $e^{-ikx}$. The right superposition is
precisely $\sin(kx+\de_0(k))$ with $\de_0(k)=\tan^{-1}\,2k$ the phase
shift in the simple model (see \eqref{eq:u}). 
To this accuracy the eigenfunction is given by
\begin{equation}
  \label{eq:u0}
  u_0(x,k)=\sin(kx+\de_0(k))\,,
\end{equation}
with a correction of order $e^{-x}$.  Of course, this is just the
first step. To cancel the $e^{-x}$ mismatch with the exact
eigenfunction one should modify the phase shift and include a $e^{-x}$
correction in the wave-function with a coefficient which is determined
by requiring that the residual correction is of order $e^{-3x/2}$.
This gives also the modified phase shift. The residual correction of
order $e^{-3x/2}$ can be eliminated by repeating the procedure.

In this way one sets up a recurrence procedure to determine both the
phase shift and the eigenfunction, as it is shown in Appendix
\ref{app:solution}. We introduce an expansion of the form
\begin{equation}\label{eq:Napprox}
  u_N(x,k) = \Imag{\exp\{-i(kx+\de_N(k))\}
\left( 1 + \sum_{n=1}^N c_n\,e^{-nx}\right)}
\end{equation}
where the coefficients $\{c_n\}$ and the phase shifts $\de_N(k)$ are
determined in such a way that the error obeys the relation
\begin{equation}
(H-E_0(k))\,u_N(x,k)= \cO{e^{-(1+\half N)\,x}} \,.
\end{equation}
In Appendix \ref{app:solution} we show that the formula at finite $N$
can be extrapolated at $N=\infty$ and we get a closed formula for the
phase shift
\begin{equation}
\label{eq:phsh}
\begin{split} 
\delta(k) &= 4\log 2\,k +\frac1{2i}
\log\left[\frac{\Gamma(1\!-\!2ik)\,\Gamma^2(1\!+\!ik)}
{\Gamma(1\!+\!2ik)\,\Gamma^2(1\!-\!ik)}\right]
\\&
={\rm arg}\left(\frac{\Gamma(1\!+\!2ik)}
{\Gamma^2\left(\half\!+\!ik\right)}\right)
\approx k\left(4\log 2-2\zeta(3)\,k^2+\cO{k^4}\right)\,.
\end{split}
\end{equation}
We also get a general expression for the approximate eigenfunctions at
all orders which can be extrapolated to $N\to\infty$ yielding the
orthonormal eigenfunctions in terms of hypergeometric functions:
\begin{equation}
  \label{eq:exact1}
  u(x,k)=\Imag{e^{i(kx+\delta(k))}\,
_2F_1\left(\half-ik,\half-ik;1-2ik;e^{-x}\right)}\,.  
\end{equation}
This representation is suited for approximating the large $x$ region,
but not for \mbox{$x\!\to\!0$} corresponding to the physical case
$\rho\!\approx\!1$ (the back-to-back dipole system for $\ee$ in the
center of mass). Notice indeed that the hypergeometric function
diverges logarithmically for \mbox{$x\!\to\!0$} but the phase shift is
precisely tuned to cancel the divergence. To see this, we use one of
Kummer's transformations for hypergeometric functions \cite{Bateman55}
to get a representation of the wave function which is manifestly
regular at the origin.  We can write Eq.~\ref{eq:exact1} in the form
\begin{equation}
  \label{eq:exact2}
  u(x,k) = \cN(k)\,e^{-ikx}\,
_2F_1\left(\half+ik,\half+ik;1;1-e^{-x}\right)\,.
\end{equation}
with $\cN(k)=\sqrt{k\pi\,\tanh k\pi}$. Notice that, in spite of the
factor $e^{-ikx}$, the eigenfunctions are real\footnote{This alternate
  presentation of the eigenfunctions was pointed out to us by V.\ A.
  Fateev.}.  This representation is well suited for small $x$
corresponding to $\rho$ close to one
\begin{equation}
  \label{eq:urho}
  u(x,k) = \cN(k)\,\left(1+(\tfrac1{4}-k^2)\,x+\cO{x^2}\right).
\end{equation}
In conclusion, for a given initial condition $I(\rho,0)$, the
$\QQ$-multiplicity is given by
\begin{equation}
\label{eq:finito}
\begin{split}
\phi(x,\tau)\!&=\!\sqrt{\frac{2}{\pi}}\int_{0}^{\infty}\,\diff{k}
  \,u(x,k)\,C(k)\,e^{-\tau\,E_0(k)}\,, \quad \\
  C(k)\!&=\!\sqrt{\frac{2}{\pi}}
  \int_0^{\infty}\diff{x}\,u(x,k)\,e^{\frac{x}{2}}\,I(e^{-x},0)\,.
\end{split}
\end{equation}
\noindent
The dispersion relation $E_0(k)$ is given by the expression
\eqref{eq:E} as in BFKL case. The initial condition at $\tau=0$ is
given in (\ref{eq:Born}). For instance, taking
$I(\rho,0)\!=\!\sqrt{\rho}\rho^{\mu}$ with $\mu>0$ we have
\begin{equation}
  C(k)=\cN(k)\frac{|\Gamma(\mu+ik)|^2}{\Gamma^2(\mu+\half)}\,.
\end{equation}
The eigenfunction expansion can be reduced to the well-known
Mehler-Fock transform \cite{Bateman55,DeAlfaro} by expressing $u(x,k)$
in terms of Legendre functions. In terms of the original density
$I(\rho,\tau)$ one then finds
\begin{equation}
  \label{eq:Legendre}
  \begin{split}
    I(\rho,\tau) &= \int_0^\infty\diff{k}\; \tilde I(k)\,
%    P_{-\half+i k}(2/\rho-1) \; e^{\tau\chi(k)}\\
    P_{-\half+i k}\left(\tfrac{2\!-\!\rho}{\rho}\right) \; e^{\tau\chi(k)}\\
    \tilde I(k) &= k\tanh\pi k
%    \int_1^\infty\diff{z}\;I\left(2/(1+z),0\right)\; P_{-\half+i k}(z)
    \int_1^\infty\diff{z}\;I\left(\tfrac{2}{1+z},0\right)\; P_{-\half+i k}(z)
  \end{split}
\end{equation}
In Fig.~1 we show, as function of $\tau$, the values of $\phi(x,\tau)$
at $x=0$ (i.e. $\rho=1$) corresponding to $\ee$ in the center of mass
system. Here we show also the corresponding $\QQ$-distribution
$I(\rho\!=\!1,\tau)$ (we have considered a simplified initial
condition).

\begin{figure}[ht] 
 \begin{center}
   \mbox{\epsfig{file=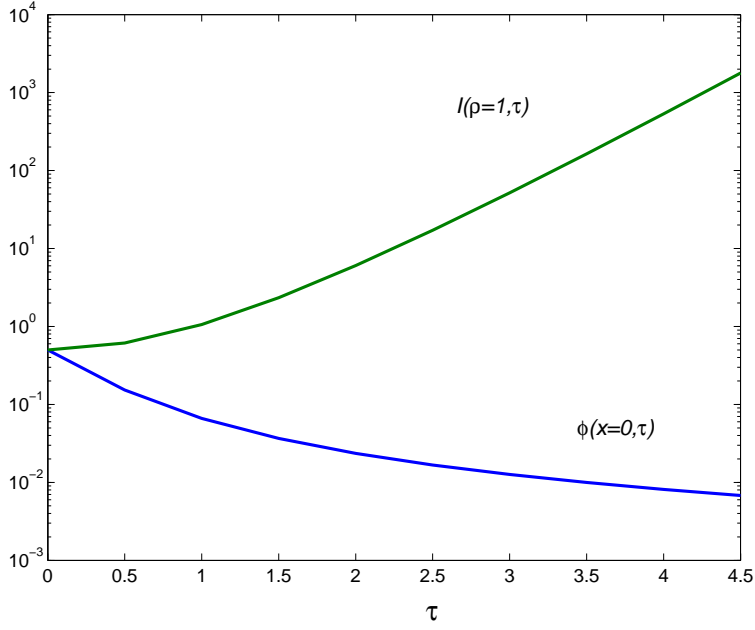,width=10.cm}}
   \caption{Evolution of $\phi$ at $\rho=1$
    obtained from \eqref{eq:finito} with initial condition
     $I(\rho,0)=\half\rho$. The corresponding $\QQ$ inclusive distribution 
$I(\rho\!=\!1,\tau)=e^{4\ln\!2\, \tau}\,\phi(0,\tau)$ is also plotted.}
 \end{center}\label{fig:phsh}
\end{figure}

In Fig.~2 we show $\phi(x,\tau)$ as function of $x$ for values of
$\tau$ which are experimentally accessible. The value at $x\!=\!0$
decreases by increasing $\tau$ as seen in Fig.~1.

\begin{figure}[ht]
  \begin{center}
    \mbox{\epsfig{file=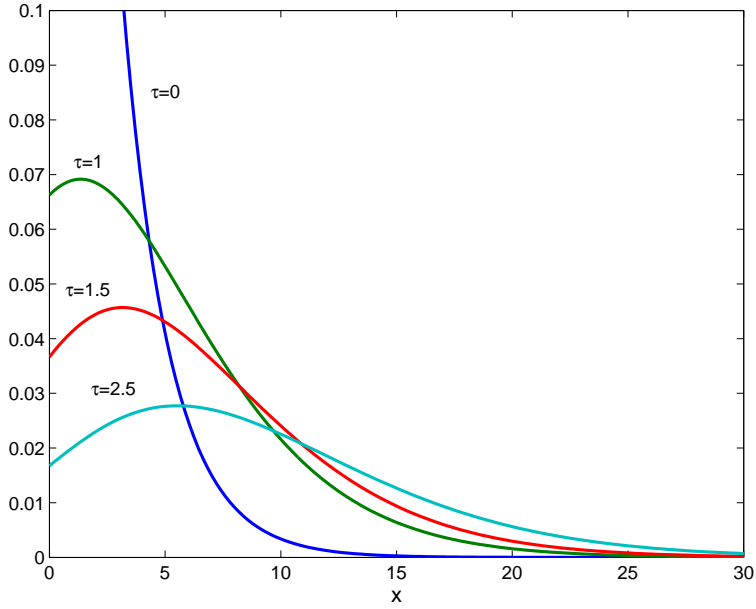,width=10.cm}}
   \caption{The distribution $\phi(x,\tau)$ obtained from 
     \eqref{eq:finito} with initial condition $I(\rho,0)=\half\rho$}.
  \end{center}\label{fig:short}
\end{figure}

From \eqref{eq:finito} one has that the solution for large $\tau$ is
dominated by the region around $k\!=\!0$.  The phase shift is a simple
analytic function of $k$ which vanishes at $k\!=\!0$.  Therefore also
the coefficient vanishes linearly with $k$.  Since the product
$u(x,k)\,C(k)$ vanishes quadratically for $k\to 0$, at large $\tau$
the distribution is given by the derivative of a Gaussian, which
implies the additional power $\tau^{-1}$ as expected (see also
\cite{MS}).  Fig.~3 shows that the Gaussian behaviour starts to set up
already at moderate values of $\tau$.

In Appendix~\ref{app:numerical} we describe a numerical approach to
obtain the solution of the $\QQ$ evolution including the large $\tau$
region, which, in practice, is faster than using the exact solution.
At large $\tau$ the shape is very accurately fitted by
\begin{equation}
\label{eq:phi-asy}
\phi(x,\tau)\approx A\,\frac{x+x_0(\tau)}
{(\tau-\tau_0)^{\frac32}}\,\exp\left\{
-\frac{(x+x_0(\tau))^2}{2D(\tau-\tau_0)}\right\}
\end{equation}
where $D=28\,\zeta(3)$ and $x_0$ is positive and weakly
$\tau$-dependent; this feature is beyond the leading saddle point
approximation.  As in the model previously discussed (see
\eqref{eq:psi-asy}) this expression vanishes at a negative value
(outside the physical region).

\begin{figure}[ht]
 \begin{center}
   \mbox{\epsfig{file=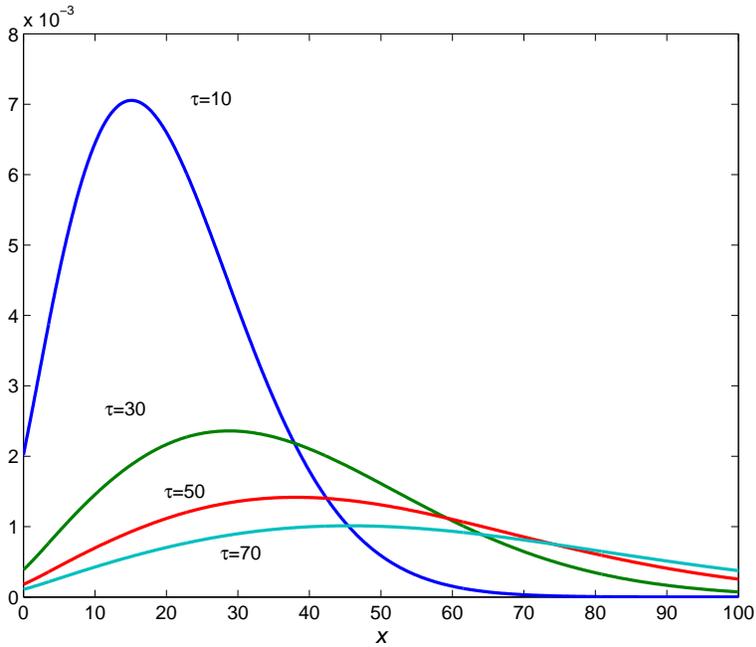,width=10.cm}}
   \caption{Long time evolution of $\phi(x,\tau)$ obtained from 
     \eqref{eq:finito} with initial condition $I(\rho,0)=\half\rho$}.
 \end{center}\label{fig:long}
\end{figure}

\section{Conclusions}\label{sec:conclusion}

The $\QQ$ and BFKL equations (see \eqref{eq:eveqMM} and
\eqref{eq:eveqBFKL}) are formally similar but originate from
multi-soft gluon distributions in two completely different phase space
regions. This fact produces important differences. First of all while
in the BFKL equation (see \eqref{eq:H-BFKL}) the variable $x$ ranges
on the entire axis, in the $\QQ$ case (see \eqref{eq:H-MM}) it is
bounded to be positive.

Moreover, due to the fact that $q_{ti}$ (the soft gluon transverse
momentum) is the argument of the QCD running coupling for each emitted
gluon, the evolution parameter $\tau$ is given by \eqref{eq:tau} in
the $\QQ$ case and by $\as\,Y$ in the BFKL case.  Since in BFKL case
$\as$ remains fixed, $\tau$ is large at high energy and then it is
physically relevant to study the asymptotic solution of the BFKL
equation. In the $\QQ$ case one has that $\tau$ involves the running
coupling at the hard scale and then increasing $Q$ one has that $\tau$
increases slowly. So the physically accessible values of $\tau$ are
not large.

A further difference between the two cases is in the physically
relevant values for the variable $x$. In the $\QQ$ case for $\ee$
annihilation in the center of mass one has $x=0$.  In the high energy
case one is interested in small impact parameter (short distance)
which corresponds to large $x$.

We have given the solution for the $\QQ$ equation which can be used in
the physical range (finite $\tau$ and small $x$). The inclusive
distribution $I(\rho,\tau)$ giving the multiplicity for a $\QQ$ pair
is
\begin{equation}
\label{eq:finito1}
I(\rho,\tau)\!=\!\sqrt{\frac{2\rho}{\pi}}\int_{0}^{\infty}\,\diff{k}
\,u(x,k)\,C(k)\,e^{\chi(k)\,\tau}\,, \qquad
\rho=\frac{1\!-\!\cos\theta}{2}\,,
\end{equation}
with $C(k)$ given in \eqref{eq:finito}, $\tau$ given in
\eqref{eq:tau}, $\chi(k)$ the BFKL characteristic function in
\eqref{eq:chi}, the eigenfunction $u(x,k)$ given in
Eq.(\ref{eq:exact1}-\ref{eq:exact2}) and $I(\rho,0)$ the Born
distribution given in \eqref{eq:Born}.  Here $\theta$ is the angle of
the dipole of partons emitting the $\QQ$ system. In $\ee$ annihilation
in the center of mass the dipole is given by the back-to-back
quark-antiquark and one has $\theta=\pi$.
In performing these integrations one may find difficulties due to the
oscillating form of the eigenfunctions.  An alternative is to use
directly the evolution equation \eqref{eq:H-MM} which can actually be
easily studied numerically. We describe in the Appendix
\ref{app:numerical} the method we used and the check of the analytical
results.

\section*{Acknowledgments} 
We thank Al Mueller for discussions on the characteristics of the
$\QQ$-multiplicity equation and for help in disentangling the
differences with the BFKL equation.  We thank Gavin Salam for pointing
out one of the most clear differences of the asymptotic behaviour in
the two cases.  We thank Vladimir A.\ Fateev and George E.\ Andrews
for generous help and for providing us a proof of
Eq.~(\ref{eq:cenne}).  We thank Roberto De Pietri for his valuable
help on some computational details.

\newpage
\appendix
\section{Analytic solution of the spectral problem}\label{app:solution} 
We start by constructing a systematic expansion as outlined in
Sec.~\ref{sec:solution}.  To this aim, it is technically more
convenient to use the original representation Eq.~(\ref{eq:eveqMM}).
Let us define
\begin{equation*}
  \phi_n \equiv \rho^{\half+n+ik}
\end{equation*}
We have
\begin{eqnarray*}
  H\,\phi_n &=& \int_0^1 \frac{\eta^{n+ik-\half}}{1-\eta}\,\diff\eta\; \phi_n 
+  \int_\rho^1 \frac{\eta^{-\half-n-ik}}{1-\eta}\,\diff\eta\;\phi_n \\
&\equiv& \mu_n(k)\,\phi_n+\nu_n(\rho,k)\,\phi_n
\end{eqnarray*}
where
\begin{eqnarray*}
\mu_n(k) &=& \int_0^1 \sum_{m\ge0}\left(\eta^{m+n+ik-\half}-\eta^m\right)\,
\diff\eta\\
&=& \sum_{m\ge0}\left(\frac{1}{m+n+\half+ik}-\frac1{m+1}\right)\\ 
&=& \psi(1)-\psi\left(\half+ik+n\right)\,.
\end{eqnarray*}
and
\begin{eqnarray}\nonumber
  \nu_n(\rho,k) &=& \int_\rho^1 \frac{\eta^{-\half-n-ik}-1}{1-\eta}\,
\diff\eta \\\nonumber
&=& \sum_{m\ge0}\left(\frac1{\half+m-n+ik}-\frac1{m+1}\right) 
-\sum_{m\ge0}\left(\frac{\rho^{\half+m-n-ik}}{\half+m-n+ik}
-\frac{\rho^{m+1}}{m+1}\right) 
\end{eqnarray}
The first sum amounts to $\psi(1)-\psi\left(\half-n-ik\right)$ which
can be added to $\mu_n$ to give
\begin{equation*}
\chi_n(k) = \mu_n+\nu_n = 
2\psi(1)-\psi\left(\half+ik+n\right)-\psi\left(\half-ik-n\right)
\end{equation*}
Note that $\chi_0$ is just Lipatov's function Eq.~(\ref{eq:E}), up to an
additive constant.  The basic relation we are to use is then the following:
\begin{equation}\label{eq:rem}
  H\,\phi_n(\rho,k) = \chi_n(k)\phi_n(\rho,k)
+ \sum_{m\ge0} \frac{\rho^{m+n+3/2+ik}}{m+1} 
- \sum_{m\ge0} \frac{\rho^{m+1}}{\half+m-n-ik}
\end{equation}
We now define 
\begin{equation}\label{eq:expN}
  \Phi_N(x,k) = \Imag{e^{-i\delta}\,
\sum_{n=0}^N\,c_n\,\phi_n(\rho,k)}\,,\;(c_0=1)
\end{equation}
which fails to be an eigenfunction for $H$ because of the various
terms originating from Eq.~(\ref{eq:rem}). The coefficients $c_n$ and
the phase shift $\delta$ will be determined by requiring that the
difference
\begin{eqnarray*}
R_N(\rho,k) &\equiv& (H-\chi_0(k))\Phi_N  \\
&=& \sum_{n\ge 1}(\chi_n(k)-\chi_0(k))\;\rho^{ik+n+\half}\\
&& + \sum_{n\ge 0}\sum_{m\ge0}  \,\frac{c_n}{m+1}\; \rho^{m+n+3/2+ik}\\
&& - \sum_{n\ge 0}\sum_{m\ge0}  \,\frac{c_n}{\half+m-n-ik}\;\rho^{m+1}
\end{eqnarray*}
be of the highest order possible $\cO{\rho^{(N+3)/2}}$. 

For the simplest possibility, $N=0$, we only have $\delta$ to be fixed
in order to cancel the leading term in $R_0$ which is
$-\rho/(\half-ik)$. This gives the same phase shift of the simplified
model of Sec.~\ref{sec:model}, namely $\delta_0(k)=\tan^{-1} 2k $.

Consider now $N=1$. We have
\begin{eqnarray*}
  R_1(\rho,k) &=& \left(1+c_1(\chi_1-\chi_0)\right)\rho^{3/2+ik} 
+ \cO{\rho^{5/2}} \\
&& -\left(\frac1{\half-ik}+\frac{c_1}{-\half-ik}\right)\,\rho + \cO{\rho^2}
\end{eqnarray*}
Now we fix $c_1$ in order to cancel the $\cO{\rho^{3/2+ik}}$ term and
the phase shift to cancel the $\cO{\rho}$ term:
\begin{equation*}
  c_1=\frac{1+2ik}{4}\,,\quad
\delta_1=\tan^{-1} 2k +\tan^{-1}\frac{2k}{3}\,.
\end{equation*}
In general, for $N$ even, we shall determine $c_1, c_2, \ldots,
c_{N/2}$ by canceling the terms of order $\cO{\rho^{3/2}},
\cO{\rho^{5/2}},\ldots,\cO{\rho^{(N+1)/2}}$ and the coefficients
$c_{N/2+1}, c_{N/2+2}, \ldots, c_{N}$ from the terms of order
$\cO{\rho^2},\ldots,\cO{\rho^{1+N/2}}$. The remaining term, the
leading one of order $\cO{\rho}$, is canceled by fixing $\delta$.
Similarly, we can proceed for $N$ odd.  The coefficients $c_n$ are
given by the recurrence relation
\begin{equation}
  \label{eq:recurrence}
c_n=\dfrac{\sum_{m=1}^n\,c_{n-m}/m}{4\sum_{m=1}^n (2m+2ik-1)^{-1}}\,,
\quad  c_0=1\,,
\end{equation}
where we have used the relation $ \chi_n(0)-\chi_n(k) =
4\sum_{j=0}^{n-1}\,(2j+1+2ik)^{-1}$.  From the first few $c_n$ it is
natural to conjecture\footnote{The proof of the conjecture has been
  found independently by G.\ E.\ Andrews and V.\ A.\ Fateev, private
  communication.}  that
\begin{equation}
  \label{eq:cenne}
  c_n=\frac{\big(\half+ik\big)_n\,^2}{n!\,(1+2ik)_n}
\end{equation}
where as usual $(a)_n=\Gamma(a+n)/\Gamma(a)$, the Pochhammer symbol.
The expression for the higher coefficients requires the solution of a
linear system and is more cumbersome.  The phase shift at the lowest
orders is given by
\begin{equation}
  \label{eq:deltak}
\begin{split}
\de_1(k)=& \tan^{-1} 2k +\tan^{-1}\frac{2k}{3} \\
\de_3(k)=& \tan^{-1} 2k -\tan^{-1} k 
  +2\tan^{-1}\frac{2k}{3} +\tan^{-1}\frac{2k}{7}\\
\de_5(k)=& \tan^{-1} 2k -\tan^{-1} k +
  \tan^{-1}\frac{2k}{3}+2\tan^{-1}\frac{2k}{5}+\tan^{-1}\frac{2k}{11} \\
\de_7(k)=& \tan^{-1} 2k -\tan^{-1} k -
  \tan^{-1} \frac{k}{2} +\tan^{-1} \frac{2k}{3}+
  2\tan^{-1} \frac{2k}{5}\\&
  +2\tan^{-1}\frac{2k}{7}+\tan^{-1}\frac{2k}{15}\\
\de_9(k)=& \tan^{-1} 2k -\tan^{-1} k -
\tan^{-1} \frac{k}{2}+\tan^{-1}\frac{2k}{3}\\&
+\tan^{-1}\frac{2k}{5}+2\tan^{-1}\frac{2k}{7}
+2\tan^{-1}\frac{2k}{9}+\tan^{-1}\frac{2k}{19}\,.
\end{split}  
\end{equation}
The regularity of the pattern allows to extrapolate to any $N$, namely
\begin{equation}
\label{eq:de-odd}
  \begin{split}
    \delta_{2n+1}(k) &= \tan^{-1} 2k  -
    \sum_{m=1}^n\,\tan^{-1} \frac{k}{m} +
    \sum_{m=1}^{2n}\,\tan^{-1}\frac{2k}{2m\!+\!1}\\&+
    \sum_{m=n+1}^{2n}\,\tan^{-1}\frac{2k}{2m\!+\!1} +
    \tan^{-1}\frac{2k}{8n+3}\,,
  \end{split}
\end{equation}
\noindent
a similar relation holding for $\delta_{2n}$.  By taking the limit
$n\to \infty$, we get the following result for the phase shift
\begin{equation}
\label{eq:exact}
\delta(k) = 4\log 2 \,k + 
\sum_{r=1}^\infty\,(-)^r\, \frac{2(2^{2r}\!-\!1)\;
\zeta(2r\!+\!1)}{2r\!+\!1}\,k^{2r+1}
\end{equation}
which is precisely the expansion of Eq.~(\ref{eq:phsh}).  In the limit
$N\to \infty$ we get the eigenfunction representation given in
\eqref{eq:exact1}.

At this stage a conservative statement is that this equation
represents just an asymptotic approximation to the solution, owing to
the fact that the higher order coefficients ($c_{N/2+1}, ..., c_N$)
grow quite rapidly with $N$. If we examine the finite sums $\Phi_N$ of
Eq.~(\ref{eq:expN}) near $x=0$ we find a typical behaviour of
asymptotic approximations, namely they oscillate around the function
$u(x,k)$ of Eq.~(\ref{eq:exact1}).

Our conjecture is that the solution in terms of hypergeometric
functions is exact, but an interchange of integration and power series
expansion is not allowed. This obstruction may be due to the fact that
the integral operator $H$ is unbounded.  From the agreement with the
numerical calculations of Appendix~\ref{app:numerical} and the overall
consistency of the whole picture, we argue that this problem can
probably be ignored, but a rigorous proof would be welcome\footnote{An
  exact solution for a class of integral equations, including the one
  considered here, and related to Tuck's equation \cite{TUCK} has been
  found in the meanwhile by V.A.Fateev (private communication, see
  \cite{DPFO}).}. 

We can cure the divergence of the finite approximation by applying
Pad\'e approximants, which effectively regulate the oscillating terms.
Notice that the hypergeometric function diverges logarithmically for
$x\to 0$ but the phase shift is precisely tuned to cancel the
divergence. To see this, we use one of Kummer's transformations for
hypergeometric functions \cite{Bateman55} to get the representation of
the wave function given in \eqref{eq:exact2}, which is manifestly
regular at the origin.  From this representation we can extract the
intercept at $\rho=1$ as a function of $k$ to be compared to the
numerical value obtained with the methods of
Appendix~\ref{app:numerical}.  This agreement gives a further support
to the consistency of the analytic solution.

\section{Numerical study}\label{app:numerical}

The evolution of the density $I(\rho,\tau)$ can be easily studied
numerically in the representation given by Eq.~(\ref{eq:H-MM}). By
introducing a normalization interval $0<x<L$ and a discrete grid
$x_0=0, x_1, x_2, \ldots, x_N=L$ with uniform spacing $a=L/N$, the
integral operator $H_0$ can be approximated by a finite $N\times N$
matrix to various degrees of accuracy. A simple choice is to adopt a
trapezoidal rule to represent the integral; the singular nature of the
kernel does not raise serious difficulties.  One can choose a
tolerance parameter $\varepsilon$, typically $10^{-8}-10^{-12}$ and
put all matrix elements less than $\varepsilon$ to zero, in such a way
that the matrix can conveniently be implemented as a {\sl sparse}
matrix.  We checked that an initial wave-function with support near
$x=L/2$ evolves according to the BFKL equation until it becomes
sensitive to the boundary.

The evolution in $\tau$ was studied using the routines provided by
{\sf matlab}.  By varying $L$ and $a$ we determined that $L=300$ and
$a=0.1$ are acceptable, in the sense that finer grids and larger $L$
do not vary the results appreciably.  Taking $\phi(x,0)=e^{-\mu x}$
one can compute $\phi(x,\tau)$, for $0<\tau<100$ in a few seconds on a
desktop computer. The snapshots in Figs.1-3 are obtained this way, but
they are indistiguishable with what one can obtain using the solution
(\ref{eq:finito}).

At large $\tau$ the shape is very accurately reproduced by
\begin{equation}
\phi(x,\tau)\approx A\,\frac{x+x_0(\tau)}
{(\tau-\tau_0)^\alpha}\,\exp\left\{
-\frac{(x+x_0(\tau))^2}{2D(\tau-\tau_0)}\right\}\,.
\end{equation}
Using $D=28\zeta(3)$ and fitting the other parameters, we find that
$\alpha$ coincides with $\frac32$ to within 0.2\%. Notice that this is
{\sl not\/} a universal property of the solutions. If the initial
shape has support at large $x$ then its evolution is effectively
described by BFKL at least until boundary effects set in.

The program can also investigate the eigenvalues and eigenfunctions.
Clearly, since we work in a finite interval, the spectrum is discrete
with spacing $\approx \pi/L$. From this preliminary numerical
investigations it emerged the nature of the eigenfunctions as {\sl
  phase--shifted standing waves\/}, and we were able to check that
{\it i\/}) the spectrum is indeed given by Eq.~(\ref{eq:chi}), and
{\it ii\/}) the phase shift vanishes for small $k$.

\end{document}